\documentclass[prd,aps,twocolumn,nofootinbib]{revtex4}

\usepackage{graphicx}

\begin{document}

\title{Early reionization by decaying particles and cosmic microwave
background radiation}

\author{S. Kasuya$^a$ and M. Kawasaki$^b$}

\affiliation{
$^a$ Department of Information Science,
     Kanagawa University, Kanagawa 259-1293, Japan\\
$^b$ Institute for Cosmic Ray Research,
     University of Tokyo, Chiba 277-8582, Japan}

\date{September, 2004}

\begin{abstract}
We study the reionization scenario in which ionizing UV photons
emitted from decaying particle, in addition to usual
contributions from stars and quasars, ionize the universe.
It is found that the scenario is consistent with both the first year
data of the Wilkinson Microwave Anisotropy Probe and the fact that the
universe is not fully ionized until $z \sim 6$ as observed by Sloan
Digital Sky Survey. Likelihood analysis revealed that rather broad
parameter space can be chosen. This scenario will be discriminated by
future observations, especially by the EE polarization power spectrum
of cosmic microwave background radiation. 
\end{abstract}


\maketitle

\setcounter{footnote}{1}
\renewcommand{\thefootnote}{\fnsymbol{footnote}}

\section{Introduction}
The dark age between recombination and today has begun to be revealed
to be very complicated by recent observations. The first witness is
the Gunn-Perterson trough observed by Sloan Digital Sky Survey (SDSS),
which implies that the universe was not fully reionized until 
$z \sim 6$ \cite{SDSS}. The second shed light on the beginning of
reionization history, which may have started as early as $z\sim 20$,
by the spectrum of cosmic microwave background (CMB) radiation
observed by Wilkinson Microwave Anisotropy Probe (WMAP) \cite{WMAP}.

Conventionally the reionization is thought to be caused by the UV photons
emitted from the early forming stars, where the ionization fraction of
the universe is approximated by a step function~\cite{Loeb}. In fact,
the early reionization may be possible with large efficiency for star
formation at high redshifts ($z\sim 20$)~\cite{FK03,Ciardi,Chen}.
(See also \cite{SYY} for difficulty in early star formation with
running spectral index suggested by WMAP.) However, it makes the
ionization fraction unity well before $z\sim 6$ if it should explain
large optical depth $\tau_{op}\sim 0.17$ suggested by WMAP. On the
contrary, the Compton optical depth for reionization is not large
enough if one considers UV photons from forming stars to be 
consistent with the data by SDSS~\cite{Ciardi,Chiu}.

In order to avoid such difficulties, one should think more
complicated reionization scenarios. For example, with the use of
unconventional initial mass function and different photon emission
processes between Population II and III stars, one can obtain the
complicated ionization history which has a period of partial
reionization before $z\sim 6$. In particular, some
authors~\cite{Cen,Hui,Sokasian} showed that the reionization occurs  
twice. 

Here we focus on another possibility. The idea is that the additional
UV photons from decaying particles can explain the earlier
reionization, while contributions from conventional stars and quasars 
make the ionization fraction unity at $z \sim 6$ \cite{KKS,HH,decay}.
Although the existence of such particles may be suspicious, the
ionization history is very realistic once we admit it. The photons
from the particle decay are characterized by three parameters:
mass, abundance, and lifetime. In this paper, we concentrate on the
case that the mass is 30 eV\footnote{
Here we consider this decaying particle as a part of cold dark matter
(CDM). It can be regarded as CDM in such a way that a scalar
condensate is coherently oscillating.}
, which emits two 15 eV photons to ionize
the universe. For the difference between larger mass and this case,
see \cite{KKS}. We will see the decaying particle scenario is
consistent with WMAP {\it and} SDSS observations.  
 
The structure of this paper is as follows. In the next section, we
review how we obtain ionization history due to photons emitted from
stars, quasars, and decaying particles. Temperature (TT) and
temperature-polarization (TE) power spectra of CMB observed by WMAP
are used to evaluate the significance of this scenario in
sect.III. Section IV is for our conclusions.

\section{Ionization history}

In our scenario, the sources for ionizing UV photons are decaying
particles, quasars and stars. Stars (and quasars) are indeed
responsible for the full reionization at $z\sim 6$. On the other
hand, contribution of UV photons from the decaying particles keeps
ionization fraction much smaller before $z\sim 6$ for a long period.

We follow the thermal history from $z>10^3$ including recombination
epoch, calculating the ionization fractions of hydrogen (HI$\!$I) 
and helium (HeI$\!$I and HeI$\!$I$\!$I), and the electron
temperature, based on the argument of Fukugita and Kawasaki
\cite{FK94}, where the hierarchical clustering scheme of the cold dark
matter scenario is used. In addition, we include the sources of
ionizing UV photons from decaying particles \cite{KKS}. Some different
treatments of stars and QSOs from \cite{FK94} can also be found in
\cite{KKS}.

We assume that the particle $\phi$ emits two photons with
monochromatic energy of half mass of that particle, i.e., 
$E_{\gamma}=m_{\phi}/2$. The number density of $\phi$-particle is
written as 
\begin{equation}
    n_{\phi} = n_{\phi}(0) (1+z)^3 e^{-\frac{t}{\tau_{\phi}}},
\end{equation}
where $\tau_{\phi}$ is the lifetime of $\phi$-particle. Notice that
those emitted photons with $E_{\gamma} > 13.6$ eV can ionize hydrogen 
atoms. Then one can write a source term for the decaying particle as
\begin{equation}
    \left(\frac{dn_{\gamma}}{dt}\right)_{dp} 
            = \frac{n_{\phi}(t)}{\tau_{\phi}}.
\end{equation}
In order to calculate how many photons are emitted, the abundance,
mass, and the lifetime of the particle should be fixed. As mentioned
above, we choose $E_{\gamma}=15$ eV, since this case represents
important features of decaying particle scenario. We calculate
ionization histories with $\tau_{\phi}=10^{14}-10^{16}$ sec (and
$10^{17}$ sec for some cases), adjusting the abundance $\Omega_{\phi}$
to get the desired value of optical depth $\tau_{op}$, whose range we
explored is $\tau_{op}=0.1-0.4$. It is defined by
\begin{equation}
    \tau_{op} = \int_{0}^{\infty} dz \sigma_T 
    \left(\frac{dt}{dz}\right)[n_e-n_e\big|_{sr}],
\end{equation}
where $\sigma_T$ is the Thomson cross section and $n_e\big|_{sr}$
is the electron number density for standard recombination. We subtract
this term in order to estimate only the effect of reionization.

\begin{figure}[!t]
\includegraphics[width=80mm]{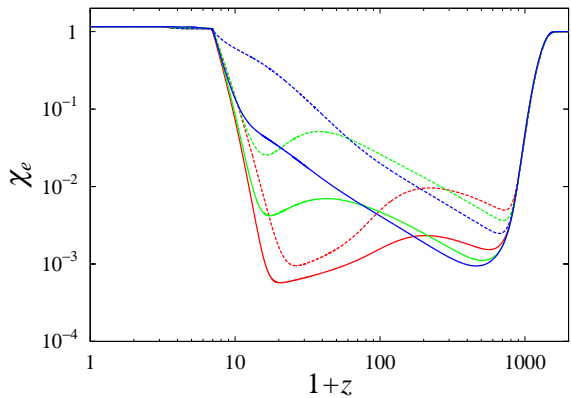}
\caption{Evolution of ionization fraction. $\tau_{\phi}=10^{14}$ sec
(red), $10^{15}$ sec (green), and $10^{16}$ sec (blue) are
plotted. For each lifetime, both  $\tau_{op}=0.1$ (solid) and 0.4
(dashed) are shown.
\label{fig_xe}}
\end{figure}

Typical ionizing histories are shown in Fig.~\ref{fig_xe}. Here
ionization fraction is defined as $\chi_e = n_e/n_H$, where $n_e$ and
$n_H$ are the number densities of electron and hydrogen atom,
respectively.  

As we will see later, cosmological parameters such as Hubble parameter,
$H_0$, matter density, $\Omega_m$, baryon density, $\Omega_b$, and
the spectral index of primordial power spectrum, $n_s$, are varied
such that they give the minimum $\chi^2$ for fitting both TT and TE
spectrum observed by WMAP in the flat $\Lambda$CDM cosmology.

As we can see in Fig.~\ref{fig_xe}, ionization due to decaying
particles endures for a long period, and the ionization fraction
remains very small ($10^{-3}\lesssim \chi_e \lesssim 10^{-1}$) until
$z\sim 6$. Even in these cases, the optical depth becomes very large.

\section{Cosmic microwave background}

Now we look for how these ionization histories are consistent with
CMB observation by WMAP. To this end, we calculate the power spectrum
with obtained evolution of the ionization fraction and matter
temperature as inputs to the code modified from CMBFAST
\cite{cmbfast}, and evaluate $\chi^2$ using the code provided by WMAP
\cite{Verde}. We search for the range $10^{14}-10^{16}$ sec, for the
lifetime of the decaying particle, and $\tau_{op}=0.1-0.4$. For fixed
lifetime and optical depth, the $\chi^2$ minimum is estimated
adjusting $\Omega_b$, $\Omega_m$, $H_0$, $n_s$, and the amplitude of
the spectrum. As an indicator for the likelihood, we use
$\Delta\chi^2=\chi^2-\chi^2\big|_{best}$ to make the contour of
likelihood with grid-based analysis. Here we take
$\chi^2\big|_{best}=1428.6$ \cite{ichikawa} as the six-parameter 
$\chi^2$ minimum for $\Lambda$CDM model with WMAP-only data.

The likelihood contours are shown in Fig.~\ref{fig_chi2} (See also the
Table~\ref{tab_chi2} below.) Here, we converted $\Delta\chi^2$ into
the confidence level for two degrees of freedom. One can see the
ionization history with $\tau_{op} \lesssim 0.2$ and
$\tau_{\phi}=10^{14}-10^{16}$ sec is consistent with WMAP data. In
general, there is little dependence on the lifetime of decaying
particle. For larger optical depth, $\tau_{\phi} \sim 10^{15}$ sec
seems more preferable a bit, although its probability is very small
because $\chi^2=1434.7$ and 1441.2 for $\tau_{op}\simeq 0.3$ and 0.4,
respectively. We omit the lifetime longer than $10^{17}$ sec from the
figure, because the ionization fraction becomes unity before 
$z\sim 6$ for $\tau_{op} \gtrsim 0.3$ even if there is no
contributions of stars and quasars. Of course, if $\tau_{op} \lesssim
0.2$, it is consistent with observations as well. For example,
$\chi^2=1432.6$ and 1430.5 for $\tau_{op}=0.2$ and 0.1, respectively. 
Notice that, for fixed optical depth, ionization histories with
lifetime longer than $10^{17}$ sec are identical to $10^{17}$ sec
case, since it is the value of $n_{\phi}/\tau_{\phi}$ that concerns
with the amount of emitted photons; i.e., the abundance should be ten
times larger for the ten times longer lifetime. This relation holds
until the abundance reaches $\Omega_{\phi} \simeq \Omega_m$.

\begin{figure}[!t]
\includegraphics[width=80mm]{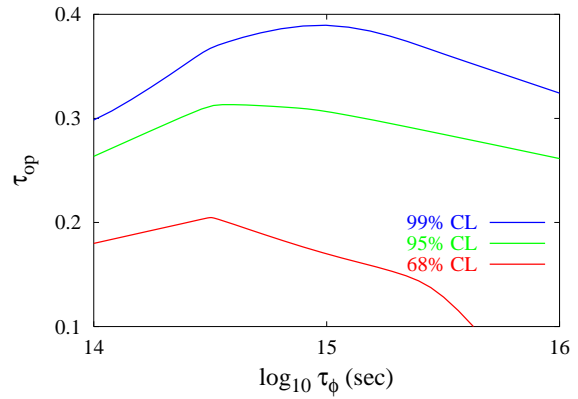}
\caption{Likelihood contour of the decaying particle scenario.
\label{fig_chi2}}
\end{figure}

\begin{table}[htbp]
\caption{Cosmological parameters which make $\chi^2$ minimum for
typical optical depth and lifetime of decaying particle. We also show
the contributions to the total $\chi^2$ from TT and TE spectra. 
Lifetime is measured in unit of sec. Baryon abundance, $\Omega_b$,
present Hubble parameter, $H_0$, and matter abundance, $\Omega_m$ are
shown the ratios to best fit values for WMAP-only data: 
$\Omega_b=0.047$, $H_0=72$ km/sec/Mpc, and $\Omega_m=0.29$. 
\label{tab_chi2}}
 \centering
   \vspace{2mm}
   \renewcommand{\arraystretch}{1.1}
   \begin{tabular}{cc|c|cc|cccc}
       \hline
       $\tau_{\phi}$ & $\tau_{op}$ & $\chi^2$ & TT & TE
       & $n_s$ & $\Omega_b$ & $H_0$ & $\Omega_m$  \\
       \hline
       $10^{14}$ & 0.417 & 1450.1 & 978.9 & 471.2 & 
       1.20 & 1.11 & 1.02 & 0.99 \\ 
       $10^{14}$ & 0.316 & 1438.9 & 976.7 & 462.1 & 
       1.13 & 1.07 & 1.01 & 0.99 \\
       $10^{14}$ & 0.203 & 1431.7 & 974.7 & 456.9 & 
       1.05 & 1.03 & 0.99 & 0.98 \\
       $10^{14}$ & 0.098 & 1429.8 & 974.7 & 455.1 & 
       0.99 & 1.00 & 0.98 & 1.00 \\
       \hline
       $10^{15}$ & 0.405 & 1438.8 & 979.7 & 459.1 & 
       1.12 & 1.08 & 1.01 & 1.00 \\
       $10^{15}$ & 0.314 & 1434.7 & 978.8 & 455.9 & 
       1.07 & 1.06 & 1.00 & 0.99 \\
       $10^{15}$ & 0.202 & 1431.5 & 976.5 & 455.0 & 
       1.02 & 1.04 & 0.98 & 1.02 \\
       $10^{15}$ & 0.097 & 1430.4 & 975.5 & 454.9 & 
       0.97 & 1.01 & 0.96 & 1.03 \\
       \hline
       $10^{16}$ & 0.407 & 1443.7 & 986.0 & 457.7 & 
       1.05 & 1.05 & 0.99 & 1.00 \\
       $10^{16}$ & 0.307 & 1437.0 & 981.6 & 455.4 & 
       1.02 & 1.02 & 0.99 & 0.97 \\
       $10^{16}$ & 0.201 & 1432.4 & 978.1 & 454.3 & 
       0.99 & 1.02 & 0.97 & 1.01 \\
       $10^{16}$ & 0.098 & 1431.4 & 975.8 & 454.9 & 
       0.98 & 1.00 & 0.98 & 1.00 \\
       \hline
       $10^{17}$ & 0.199 & 1432.6 & 978.6 & 454.1 & 
       0.98 & 1.03 & 0.96 & 1.03 \\
       $10^{17}$ & 0.099 & 1430.5 & 976.0 & 454.6 & 
       0.96 & 1.02 & 0.95 & 1.05 \\
       \hline
   \end{tabular}
\end{table}

We should mention that $\chi^2$ itself is smaller if reionization
takes place instantaneously for the same optical depth. As we have
mentioned several times, however, such instantaneous reionization
histories do not respect the Gunn-Peterson trough at $z\sim 6$
observed by SDSS. On the other hand, optical depth becomes 
$\simeq 0.04$ if the ionization fraction at $z\sim 6$ is suppressed
enough to meet SDSS data. (When using our code without contribution of
decaying particle, we get $\tau_{op}\simeq 0.05$.)

\begin{figure}[!t]
\includegraphics[width=80mm]{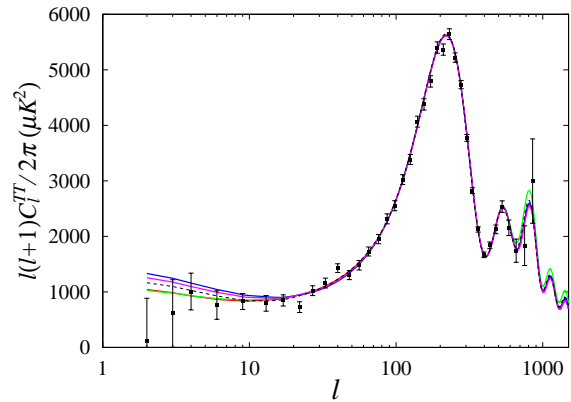}
\caption{TT power spectra for typical ionization histories: 
$\tau_{\phi}=10^{14}$ sec and $\tau_{op}\simeq 0.1$ (red),
$\tau_{\phi}=10^{15}$ sec and $\tau_{op}\simeq 0.4$ (green),
$\tau_{\phi}=10^{16}$ sec and $\tau_{op}\simeq 0.3$ (blue),
$\tau_{\phi}=10^{17}$ sec and $\tau_{op}\simeq 0.2$ (purple),
and instantaneous reionization which leads to $\tau_{op}\simeq 0.17$
(dashed black). In addition, WMAP data is also shown.
\label{fig_tt}}
\end{figure}

\begin{figure}[!t]
\includegraphics[width=80mm]{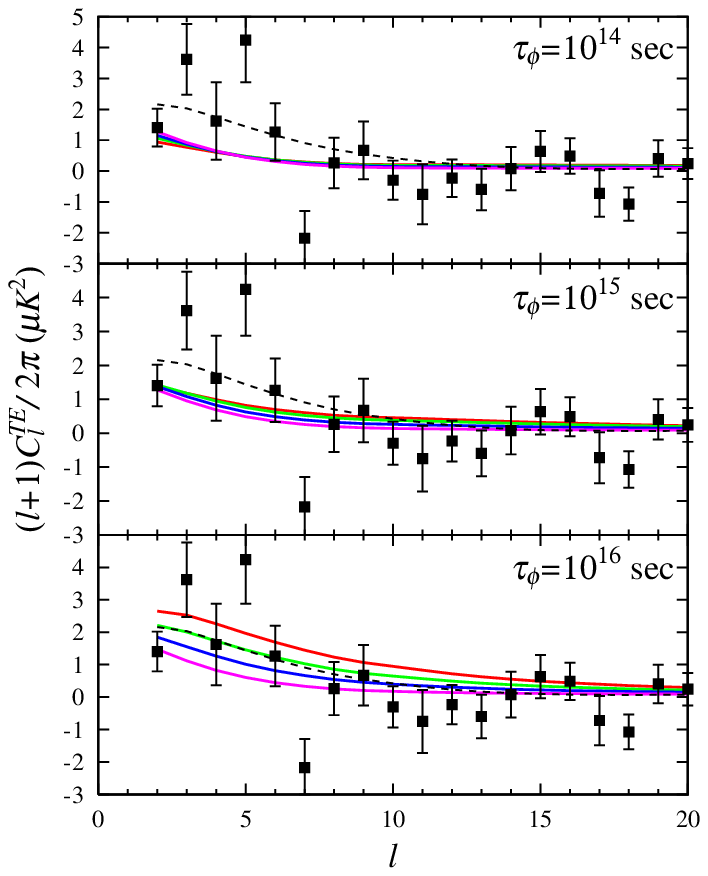}
\caption{TE power spectra for typical ionization histories for
$\tau_{\phi}=10^{14}$ sec (top panel), $10^{15}$ sec (middle panel),
and $10^{16}$ sec (bottom panel). In each panel, the lines are for
$\tau_{op}\simeq 0.4$ (red), 0.3 (green), 0.2 (blue), and 0.1
(purple). The instantaneous reionization with $\tau_{op}\simeq 0.17$
is shown in dashed black line, and the data is from WMAP.  
\label{fig_te}}
\end{figure}

\begin{figure}[!t]
\includegraphics[width=80mm]{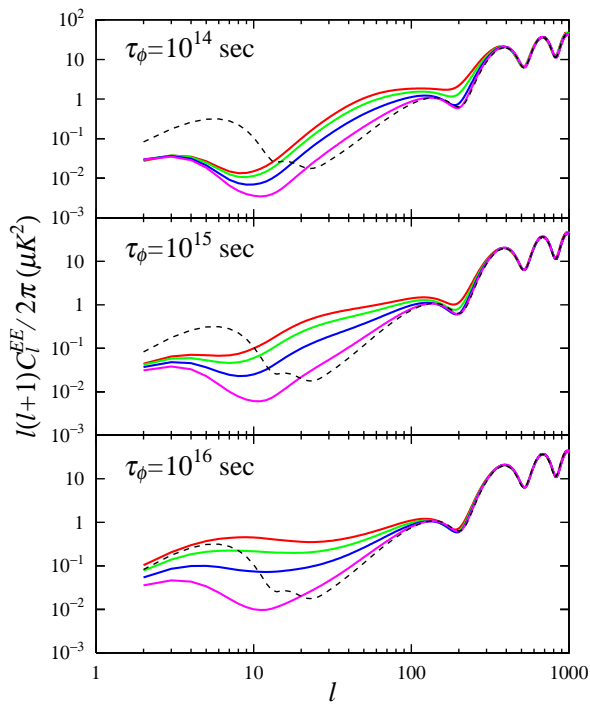}
\caption{EE power spectra for typical ionization histories for
$\tau_{\phi}=10^{14}$ sec (top panel), $10^{15}$ sec (middle panel),
and $10^{16}$ sec (bottom panel). In each panel, the lines are for
$\tau_{op}\simeq 0.4$ (red), 0.3 (green), 0.2 (blue), and 0.1
(purple). The instantaneous reionization with $\tau_{op}\simeq 0.17$
is shown in dashed black line.
\label{fig_ee}}
\end{figure}

Now let us see the CMB spectra in detail.
We show TT, TE, and EE power spectra of CMB in
Figs.~\ref{fig_tt}$-$\ref{fig_ee}. For TT and TE spectra, we also plot
the WMAP data. One can see from Fig.~\ref{fig_tt} that all the models
look almost identical for $\ell\gtrsim 100$. The biggest differences
can be seen in low multipole region. This happens because large
optical depth reduces height of the acoustic peaks, which results in
larger power in low multipoles ($\ell \lesssim 100$) when the height
of the first acoustic peak is fixed. In addition to this, the spectrum
should be more tilted to the blue side for larger optical depth in
order to fit the data, i.e., the well-known $\tau_{op}-n_s$
degeneracy, which lowers the plateau. Thus the outputs are made by
competition between the two opposite effects.  

Important point at present is that TE power spectrum does not
have much ability of determining ionization histories using the first
year WMAP data. The errors are still large, and even though those
histories which do not seem to fit the TE data well have small
$\chi^2$ (see Fig.~\ref{fig_te} and Table~\ref{tab_chi2}) if the fit
to TT data is good enough. This may lead to the fact that there is
little dependence of the $\chi^2$ on the lifetime especially for small
optical depth region. Reversely, we could distinguish among different
histories with various optical depth so long as the errors diminish.

Although one may discriminate among the models using TE spectrum, EE
spectrum seems the most promising tool for that purpose. Most
different feature of decaying particle scenario is the larger power at
$\ell \sim$(a few)$\times 10$ due to enduring UV photon emission. 
Moreover,  we may distinguish even the lifetime of decaying particle
depending on whether there is a dip around $\ell \sim 10$ and the
steepness of the slope below $\ell \sim 100$. Therefore, EE spectrum
expected to be seen by Planck satellite or even WMAP should (dis)prove
the photons from decaying particles in the near future.

\section{Conclusion}
We have sought for a natural reionization history which respects both
WMAP and SDSS observations. Decaying particles provide UV photons
which reionize the universe from rather early period as suggested by
WMAP, but keeps the ionization fraction very small until usually
considered UV sources such as stars and quasars bring up the fraction
to unity abruptly at $z\sim 6$ as observed by SDSS. The central concern
is how well such scenario goes on. To this end, we have calculated TT
and TE power spectra of CMB, and done the likelihood analysis. We
have found decaying particle scenario is consistent with both
observations in rather broad parameter space. Even particles with
lifetime $10^{14}$ sec, corresponding to the redshift $z\sim 280$,
which have the abundance to give $\tau_{op}\lesssim 0.3$, seems good
as well. 

Looking into the detail, it is the TT power spectrum that determines
the value of $\chi^2$ dominantly. This is the reason why the
reionization history, whose TE spectrum does not seem to fit the data
so well, can have small $\chi^2$, such as 1429.8 for 
$\tau_{op}\simeq 0.1$ and $\tau_{\phi}=10^{14}$ sec, for example. When
the quality of TE data improves, then one may be able to tell which
reionization history is preferable. As for the ability of
discriminating each history, EE spectrum may be the best tool. 
Decaying particle scenario has a peculiar feature that there is much
more excess of the power in the range $20 \lesssim \ell \lesssim 100$
compared with instantaneous reionization case. Moreover, it may be
possible to observe even the lifetime of the decaying particle. 

Finally, we comment on a maybe serious problem if one regards early 
forming stars as the whole source of ionization photons because of the
less powers on smaller scales observed by, say, 2dfGRS and SDSS. Since
the decaying particle scenario does not owe to early forming stars as
the UV source, it is completely free from such a problem. We have
checked that our model works also for the running spectral
index suggested by WMAP. Since all the astrophysical UV sources may
suffer this problem, the decaying particle could be the better source
of ionizing UV photons.

\section*{Acknowledgments}
SK is grateful to K. Ichikawa for useful discussions.
MK is supported by Japanese Grant-in-Aid for Science Research Fund of
the Ministry of Education, No.14540245 and No.14102004.



\end{document}